\documentclass[preprint,showpacs,preprintnumbers,amsmath, amssymb, 11pt]{revtex4}
\usepackage{amsmath,amssymb,graphicx}
\usepackage{epstopdf}
\usepackage {amssymb}
\newcommand{\nc}{\newcommand}
\nc{\ba}{\begin{eqnarray}}
\nc{\ea}{\end{eqnarray}}
\newcommand\be{\begin{equation}}
\newcommand\ee{\end{equation}}

\nc{\x}{{\bf{x}}} \nc{\y}{{\bf{y}}} \nc{\f}{{\bf{f}}}
\nc{\vo}{{\bf{v}}} \nc{\p}{{\bf{p}}} \nc{\dep}{{\delta\phi}}
\nc{\D}{\overline{\mbox{D3}}}

\input{epsf.sty}

\preprint{IPM/P-2010/047  }
\begin{document}

\title{ Fields Annihilation and Particles Creation in DBI inflation}

\author{Hassan Firouzjahi$^{1}$}
\email{firouz(AT)ipm.ir}
\author{Salomeh Khoeini-Moghaddam$^{2,1}$}
\email{khoeini(AT)ipm.ir}

\affiliation{$^{1}$ School of Physics, Institute for Research in
Fundamental Sciences (IPM), P. O. Box 19395-5531, Tehran, Iran}
\affiliation{$^{2}$ Department of Physics, Faculty of
Science,Tarbiat Mo'allem university, Tehran, Iran }
\date{\today}
\begin{abstract}
We consider a model of DBI inflation where two stacks of mobile branes are moving ultra relativistically in a warped throat. The stack closer to the tip of the throat is annihilated with the background anti-branes while inflation proceeds by the second stack.  The effects of branes annihilation and particles creation during DBI inflation on the curvature perturbations
power spectrum and the scalar spectral index are studied.
 We show that for super-horizon scales, modes which are outside the sound horizon at the time of branes collision, the spectral index has a shift to blue spectrum compared to the standard DBI inflation. For small scales  the power spectrum approaches to its background DBI inflation value with the decaying superimposed oscillatory modulations.

\end{abstract}
\maketitle

\section{Introduction}

Inflation proved to be very successful as a theory of early universe and structure formation \cite{Guth:1980zm} which is strongly supported by cosmic observations \cite{Komatsu:2010fb}. The simplest models of inflation are based on a scalar field minimally coupled to gravity with a potential flat enough to support an extended period of inflation. Despite its observational successes there is no deep theoretical understanding of inflation and the nature of the inflation field. There have been many attempts during past decade to embed inflation within the context of string theory, for a review see  \cite{HenryTye:2006uv,Cline:2006hu,Burgess:2007pz,McAllister:2007bg,Baumann:2009ni, Mazumdar:2010sa}
and references therein. 

Brane inflation is an interesting realization of inflation from string theory \cite{dvali-tye,Alexander:2001ks,collection,Dvali:2001fw}. In its original form, it
contained a pair of D3  and anti D3 branes moving in the Calabi-Yau (CY) compactification. The inflaton field is the radial distance between them so in this sense the inflaton field has a geometric interpretation in string theory. Inflation ends when the distance between the brane and the anti-brane reaches the string length scale where a tachyon develops in the open strings spectrum stretched between the pair. Inflation ends soon after tachyon formation and the energy stored in branes tensions are released into closed string modes  \cite{HenryTye:2006uv}. However, it was soon realized that the 
potential between the pair of brane and anti-brane is too steep to allow a long enough period of 
slow-roll inflation. To flatten the potential, it was suggested to put the pair of brane and anti-brane  inside a warped throat \cite{Klebanov:2000hb, Giddings:2001yu, Dasgupta:1999ss}, where the potential between D3 and $\D$ is warped down as in Randall-Sundrum scenario
\cite{Randall:1999ee, Kachru:2003sx, Firouzjahi:2003zy, Burgess:2004kv, Buchel, Iizuka:2004ct, Firouzjahi:2005dh, Baumann:2006th, Burgess:2006cb, Baumann:2007ah, Chen:2008au, Cline:2009pu}.

As a novel feature of brane inflation the question of branes annihilation and particles creation during inflation was studied in \cite{Battefeld:2010rf}. In this picture two stacks of branes, located at different places inside the throat,  are moving in the slow-roll limit towards the bottom of the throat where anti-branes are located.  The stack closer to the tip will be annihilated during inflation transferring its energy into closed string modes. The second stage of inflation is driven by the remaining stack of branes until it is annihilated by the remaining anti-branes at the bottom of the throat. The process of fields annihilation \cite{Battefeld:2008py, Battefeld:2008ur} and particle creations \cite{Barnaby:2009mc, Barnaby:2009dd, Barnaby:2010ke,Barnaby:2010sq, Green:2009ds, Brax:2010tq, Cai:2010wt} can have interesting observational consequences on CMB.  Here we generalize the results in \cite{Battefeld:2010rf} to the case where the mobile stacks of branes are moving ultra relativistically as in DBI inflation \cite{Alishahiha:2004eh}. As in \cite{Battefeld:2010rf}, the stack closer to the tip is annihilated                by the background anti-branes resulting in particles creation during inflation while the second stage of inflation is driven by the remaining stack. 

The rest of the paper is organized as follows. In section \ref{setup} we present our set up and background solutions for DBI inflation. In section \ref{perturbation} we study the curvature perturbations  and present our matching conditions. In section
\ref{power} we obtain the power spectrum of curvature perturbations and the spectral index.
Discussions and conclusions are presented in section \ref{conclusion}.

\section{The model and background equations}
\label{setup}
\begin{figure}[t]
   \centering
  \includegraphics[width=2in,angle=0]{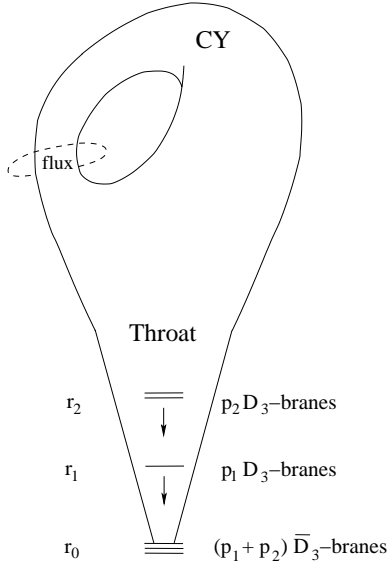}
\caption{A schematic view of the set up. There are two stacks containing $p_1$ and $p_2$
coincident branes inside the throat. The stacks are moving relativistically towards the bottom of the throat where there are $p_1+ p_2$ anti-branes. The figure is borrowed from \cite{Battefeld:2010rf}
}
\vspace{0.5cm}
\label{throat1}
\end{figure}

\begin{figure}[t]
   \centering
  \includegraphics[width=4.5in,angle=0]{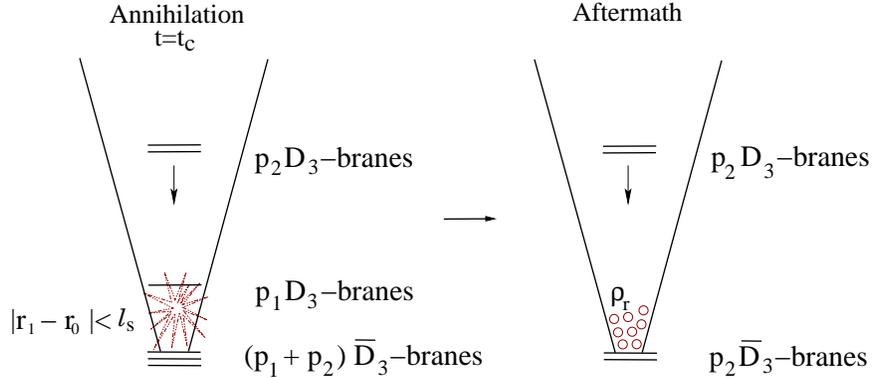}
\caption{A schematic view of the annihilation event at $t_c$. The stack closer to the tip of the throat is annihilated during inflation transferring its energy into radiation. The second stage of inflation is driven by the remaining stack containing $p_2$ branes. Inflation ends when this stack is annihilated by the remaining $p_2$ anti-branes at the bottom of the throat. The figure is borrowed from \cite{Battefeld:2010rf}.
}
\vspace{0.8cm}
\label{throat2}
\end{figure}

In this section we present our set up. As explained above we consider two stacks containing $p_1$ and $p_2$ coincident branes. The positions of the stacks are represented by functions 
$r_1(t)$ and $r_2(t)$ inside the throat. There are $p_1+p_2$ anti-branes located at the bottom of the throat, $r_0$.  We assume that the first stack is closer to the tip of the throat so it is annihilated during inflation transferring its energy into closed strings modes. The second stage of inflation is driven by the remaining stack of $p_2$ branes. For a schematic view, see {\bf Fig.} \ref{throat1} and {\bf Fig.} \ref{throat2}.


As usual the metric of the throat is 
 \ba\label{metric}
 ds^2=h^{-\frac{1}{2}}(r)g_{\mu\nu}dx^\mu dx^\nu+h^{\frac{1}{2}}(r)(dr^2+r^2d\Omega_5^2) \, ,
\ea
where $r$ is the radial coordinate,  $d \Omega_5^2$ represents the internal five-dimensional azimuthal directions which we may take to be an $S^5$ and  $h(r)$ is the warp factor
\ba
\label{warp-factor}
 h(r)=\frac{L^4}{r^4}\, .
\ea
Here $L$ is the AdS length scale of the throat which is created by $N$ coincident background 
 branes located at $r=0$ (for a review see \cite{Herzog:2001xk}),
 \ba  
 \label{L}
 L^4 =4\pi g_s N \alpha'^2 \, ,     
 \ea
where $g_s$ is the perturbative string coupling and $\alpha'= l_s^2$ where $l_s$ is the string length scale. The mobile stacks of $p_1$ and $p_2$ branes are located at positions $r_1$ and 
$r_2$ as probe branes.  In order for our probe branes approximation to be valid, we need $p_1, p_2 \ll N$. In addition there are $p_1+ p_2$ $\D$ branes at the bottom of the throat $r_0$. Once the distance between the stack of $p_1$ branes and these anti-branes reaches at the order of $l_s$ the open strings stretched between them become tachyonic, indicating an instability in the system. Shortly after tachyon formation, the stack of $p_1$ branes is annihilated by $p_1$
anti-branes at the bottom of the throat. Consequently, the energy stored in $p_1$ branes and anti-branes tension, which is $2 p_1 T_3 h(r_0)^{-2}$, is transferred into closed strings modes.
For simplicity, we assume that these closed strings modes are in the form of massless particles (radiation). Of course a combinations of massive and massless particles would be created. However, once these particles are created they will be diluted quickly by the background inflation so for our practical purposes, there are not much differences between massive and massless particle creations. 

The goal of this work is to examine the effects of  branes annihilation and particles creation on the remaining field which drives the second stage of inflation. In  \cite{Battefeld:2010rf} this was studied for the case where the stacks are moving slowly. Here we generalize those analysis to our case at hand where the stacks are moving ultra relativistically \cite{Alishahiha:2004eh, Chen:2004gc, Shandera:2006ax, Bean:2007eh, Bean:2007hc, Chen:2006nt}.

The Dirac-Born-Infeld (DBI) action for the stacks of $p_1$ and $p_2$ branes moving relativistically  in the background of Eq. (\ref{metric}) is 
\ba
\label{action}
S=\sum_{I=1}^2 p_I
 T_3\int\sqrt{-g} \, d^4x \left[h^{-1}(r_I) \left( 1-\sqrt{1-h(r_I) \, {\dot{r}_I}^2}\right)-V_I(r_I)\right] \, ,
\ea
where $T_3$ is the D3-brane tension with $T_3 L^4=\frac{N}{2\pi^2}$. We also have added the unknown contributions $V_1(r_1)$ and $V_2(r_2)$ from the back-reactions of the 
background fluxes, branes and Kahler modulus stabilization 
on the dynamics of mobile branes \cite{Kachru:2003sx, Firouzjahi:2005dh}. As usual, we shall continue from the phenomenological point of views and parametrize these potentials to be suitable for inflation
\cite{Alishahiha:2004eh, Chen:2004gc, Shandera:2006ax, Bean:2007eh, Bean:2007hc, Chen:2006nt, Thomas:2007sj, Huston:2008ku}.
 
To proceed further, we simplify the DBI action into the following form which is more suitable for our analysis 
 \ba
 \label{action2}
 S=\sum_{I=1}^2\int d^4x\sqrt{-g}\left[ f_I^{-1}\left(1-\sqrt{1-f_I{\dot{\phi}_I}^2}\right)-V_I(\phi_I)
 \right] \, ,
 \ea
where  $\phi_I\equiv\sqrt{T_3p_I}r_I$,  $f_I \equiv \frac{\lambda_I}{\phi_I^4}$ and 
 $\lambda_I\equiv T_3L^4p_I    = \frac{Np_I}{2\pi^2}$.  This form of multiple field DBI inflation has been studied extensively in the past \cite{Cai:2008if}.

Now we are ready to promote the system into a cosmological set up. The background four-dimensional FRW metric is 
\ba
ds^2 = -dt^2 + a(t)^2d\vec{x}^2 \, ,
\ea
where $a(t)$ is the scale factor. The Friedmann equation and the energy conservation equations are
\ba  
\label{back1}
3H^2=\frac{\rho}{M^2_p} \quad \quad  ,\quad \quad 
\dot{\rho}+3H(\rho+P)  =0 \, ,
\ea
where $H= \dot a/a$ is the Hubble expansion rate. The energy density $\rho$ and the pressure $p$ are given by

\ba
\rho=\sum_{I} \left[f_I^{-1}(\gamma_I-1)+V_I \right] \quad  \quad , \quad \quad 
p=\sum_{I}\left[f_I^{-1}(1-\gamma_I^{-1})-V_I \right] \, .
\ea
Here the Lorentz factor for each field, $\gamma_I$, is defined by
 \ba
 \gamma_I=\frac{1}{\sqrt{1-f_I\dot{\phi}_I^2}} \, .
 \ea
Finally, the background Klein-Gordon equation for each field is 
\ba
\label{eqm phi}
\ddot{\phi}_I+3H\gamma_I^{-2}\dot{\phi}_I+\frac{3}{2}\frac{{f'_I}}{f_I}\dot{\phi}_I^2-\frac{{f'_I}}{f_I^2}+\gamma_I^{-3} \left({V_I'}+\frac{{f'_I}}{f_I^2} \right)=0 \, .
\ea

So far we have not specified the form of the potentials, $V_I(\phi_I)$. As explained above, these potentials arise from the back-reactions of the fluxes, branes and Kahler modulus stabilization  to the mobile branes. It is a non-trivial question in string theory as how one can calculate these back-reactions  \cite{Baumann:2006th, Burgess:2006cb, Chen:2008au} concretely. However, to be specific, we consider the phenomenological approach where the potentials have the form $V_I = \frac{m_I^2}{2} \phi_I^2$. We shall also keep the masses $m_I$ undetermined  except that their magnitudes should be such that they give the desired number of e-foldings and fit the CMB observations such as the COBE normalization.

We are interested in the limit where the branes are moving ultra relativistically inside the throat so one can not expand the square root in Eq. (\ref{action2}) perturbatively in terms of $\dot \phi$. This corresponds to the case where $\gamma \gg1$  and the stacks of branes are approaching the speed limit $\dot \phi_I^2  \simeq f_I^{-1} $.  In this limit Eq. (\ref{eqm phi}) reduces to
\ba
\label{eqm phi app}
\ddot{\phi}_I+\frac{3}{2}\frac{{f'_I}}{f_I}\dot{\phi}_I^2-\frac{{f'_I}}{f_I^2}=0 \, ,
\ea
which  can be verified to have the speed limit solution $\phi_I(t) \simeq \frac{\sqrt \lambda_I}{t}$. One can plug this solution back into Eq. (\ref{eqm phi}) to find the next leading correction and
\ba
\label{phi-back}
\phi_I(t)  \simeq  \frac{\sqrt \lambda_I}{t} \left(  1- \frac{9 H^2}{2 m_I^4 \, t^2}
\right) \, .
\ea
To get this solution, we have neglected the variation of $H$ during inflation to leading order.

To obtain the solution for the scale factor $a(t)$ we require that the energy density is dominated by the potential energy so the Friedmann equation (\ref{back1}) reduces to $\dot a/a \propto 1/t$.  Below we find the background solutions for both inflationary stages, before the collision and after the collision.

\subsubsection{Before Collision}

Before collision both stacks contribute to the energy density and the Friedmann equation can easily be solved to give
\ba
\label{dot-a}
 \frac{\dot{a}}{a}=\frac{{\Lambda_{-}}}{t} \, ,
 \ea
where the dimensionless number $\Lambda_-$ is defined by
\ba
\label{lambda-}\Lambda_{-}\equiv \left( \sum_{i}\frac{\lambda_i m_i^2}{6 M_p^2}  \right)^{1/2} \, .
\ea
Working with the number of e-foldings $N(t)$ as the clock, $d N = H dt$, one obtains
\ba
\label{N}
N(t)= \Lambda_- \ln \frac{t}{t_{in}} \, ,
\ea
where $t_{in}$ is the time at the start of inflation when $N=0$. Define $N=N_c$ the time when the first branes collision take place and the stack of $p_1$ branes is annihilated by $p_1$ background anti-branes. Similarly, define 
$\phi_I = \phi_{I c}$ as the corresponding fields values at the time of collision. Combining
Eqs. (\ref{N}) and (\ref{phi-back}) yield
\ba
\label{phiIc}
\phi_{Ic} \simeq \phi_{I in} e^{- \frac{N_c}{\Lambda_-}} \, ,
\ea
where $\phi_{I in}$ are the corresponding fields values at the start of inflation.

One can find an approximate formula for the time of branes collision as follows. As mentioned before, the time of branes annihilation is when the physical distance between the stack of 
$p_1$ branes  and anti-branes located at the bottom of the throat $r_0$ reaches the string length scale $l_s$. Starting with the metric (\ref{metric}) the physical distance between the first stack and the anti-branes, $\Delta \ell$, 
is calculated to be
\ba
\Delta \ell = \int_{r_0}^{r_{1}} h(r)^{1/4} dr = L  \, \ln \frac{r_1}{r_0} \, .
\ea
Setting $\Delta \ell= l_s$  and using Eq. (\ref{phiIc}) the onset of branes collision, $N=N_c$, is obtained to be
\ba
\label{Nc}
N_c &\simeq& \Lambda_- \left( \ln \frac{\phi_{1 in}}{\phi_{1A}} - \frac{l_s}{L} \right) \nonumber\\
&\simeq& \Lambda_- \left( \ln \frac{\phi_{1 in}}{\phi_{1A}} - (4 \pi g_s N)^{-1/4} \right) \, ,
\ea
where to get the final relation Eq. (\ref{L}) has been used. Here we defined $\phi_{IA} \equiv \sqrt{p_I T_3 } r_0  $ as the values of the corresponding fields at the position of the anti-branes. 

Correspondingly, the position of $\phi_2$ at the time of collision is obtained to be 
\ba
\phi_{2c} \simeq \frac{\sqrt{\lambda_2}}{t_c} 
\simeq \phi_{2 in} e^{- \frac{N_c}{\Lambda_-}} \, .
\ea

Also it is instructive to calculate $\gamma_I$. Using Eqs. (\ref{phi-back}) and (\ref{dot-a})
in the expression of $\gamma_I$ one finds
\ba
\gamma_{I }^- \simeq  \frac{ m_I^2 \, t^2}{3  {\Lambda_-}} \, .
\ea
This indicates that as inflation proceeds, $\gamma_I$ increases like $t^2$.

\subsubsection{After Collision}
After the first stack is annihilated, the second stage of inflation is driven by the remaining stack
of $p_2$ branes and the scale factor is given by $\frac{\dot a}{a} = \frac{{\Lambda_+}}{t}$ where now 
\ba
\label{Lambda+}
\Lambda_+ \equiv  \left( \frac{\lambda_2 m_2^2}{6 M_P^2}\right)^{1/2} \, .
\ea 
Defining the total number of e-foldings by $N_T$, with $N_T \sim 60$ to solve the flatness and the horizon problem, one has
\ba
N_T - N_c \simeq {\Lambda_+} \ln \frac{t_f}{t_c} \, ,
\ea
where $t_f$ is the time of end of inflation when the second stack is annihilated by the remaining 
$p_2$ anti-branes. As before, this happens when the physical distance between the second stack and the anti-branes reaches the string scale $l_s$ which can be used to determine the value of $\phi_2$ at the end of inflation  $\phi_{2f} \simeq \phi_{2A}  \exp( l_s/L)$. Similarly, one can show
\ba
\label{NT}
N_T &\simeq& N_c + {\Lambda_+} \ln   \frac{\phi_{2 c}}{\phi_{2 f}} \nonumber\\
&\simeq& 
\left(  1- {\frac{\Lambda_+}{\Lambda_-} } \right) N_c + 
{\Lambda_+}  \ln \left[  \frac{\phi_{2 in}}{\phi_{2 A}} - (4 \pi g_s N)^{-1/4} \right] \, ,
\ea 
One should arrange the background parameters such as $g_sN$, $\Lambda_{\pm}$
and  $\phi_{I in}$ such that large enough $N_T$ can be obtained to solve the flatness and the horizon problem.

Correspondingly, the Lorentz factor after the collision is 
\ba
\gamma_2^+ \simeq    \sqrt{ \frac{2}{3}}  \frac{ m_2 M_P t^2}{ \sqrt {\lambda_2}} \, .
\ea

One can find a measure of the magnitudes of $\Lambda_{\pm}$ as follows. For DBI inflation to match the COBE normalization, one requires a large background charges such that $g_s N \gg1$ \cite{Alishahiha:2004eh, Baumann:2006cd} so one can safely neglect this term in Eqs. (\ref{NT}) and (\ref{Nc}). Assuming that there is not exponential hierarchy between $\phi_{2 in}$ and $\phi_{2 A}$ one concludes that
${\Lambda_+} \lesssim N_T$. However, by keeping the ratio $ \frac{\phi_{2 in}}{\phi_{2 A}}$ exponentially large in the light of \cite{Giddings:2001yu}, one can lower ${\Lambda_{+}}$ by a factor of ten or so. In our analysis below, we shall take take ${\Lambda_{+} }    \simeq   {\Lambda_{-} } \gg 1$.

As shown in \cite{Alishahiha:2004eh} significant amount of non-Gaussianities can be created in DBI inflation where the branes are moving ultra-relativistically. The non-Gaussianity parameter $f_{NL}$ is related to the Lorentz factor via $f_{NL}  \simeq  - 0.3 \gamma^2$. To satisfy the  WMAP constraints on $f_{NL}$ one requires that $\gamma < 31$   \cite{Shandera:2006ax, Bean:2007eh, Bean:2007hc}.

\section{Perturbations}
\label{perturbation}

Having studied our inflationary background in some details, now we consider the perturbations in this background. As discussed in \cite{Battefeld:2010rf} a complete treatment of perturbations in our set up with branes annihilation and particles creation is a non-trivial task.
The process of field annihilation and particles creation are determined by the dynamics of open string tachyon condensation. In the presence of background branes and fluxes this is a non-trivial phenomena \cite{Jones:2003ae, Jones:2002sia}. We shall instead take the phenomenological approach and impose some physically reasonable approximations to bypass these difficulties. 

Before the collision we have two scalar fields $\phi_1$ and $\phi_2$ corresponding to two stacks of branes moving relativistically. After the collision, the energy associated with field $\phi_1$ is transferred into closed strings modes, which we approximate them as massless particles. The process of field annihilation and closed strings formation takes some time scale, $\Delta t$. In our set up where the collision happens at the bottom of the throat $\Delta t$ is given by the inverse of the warped string mass scale, that is  
$ \Delta t \simeq h(r_0)^{1/4} m_s^{-1} $ where
$m_s = l_s^{-1}$ is the string theory mass scale. Note that the factor $h(r_0)^{1/4}$ in 
$\Delta t$ is in the light of Randall-Sundrum idea \cite{Randall:1999ee}.  In our effective four-dimensional approach where $H \ll m_s$, we can take the process of field annihilation and particles creation as instantaneous and set $\Delta t \sim 0$ for practical purposes. This approximations is certainly true for large scale perturbations. 

To be slightly more specific, we can borrow the results of \cite{Barnaby:2004gg, Kofman:2005yz, Firouzjahi:2005qs, Chen:2006ni}. As mentioned before, after brane and anti-brane annihilation their energy  is transferred into closed strings modes. Subsequently, 
these modes cascade down to graviton zero mode and the Kaluza-Klein (KK) excitations. 
This happens in a time scale controlled by the inverse of the warped sting mass scale, much shorter than $H^{-1}$. Subsequently, the massive KK modes decay into lighter modes.
Because of the warp factor, the wave function of these modes are highly peaked at the bottom 
of the throat where the annihilation takes place.
Their effects on the remaining stack of $p_2$ branes at position $r_2$ may be negligible in the limit where the stack of $p_2$ branes is far away from the tip of the throat.

Even in the instantaneous brane annihilation approximation it is not clear within string theory set up as how to turn off field $\phi_1$ and turn on radiation. To bypass this shortcoming we follow the practical approach employed in \cite{Battefeld:2010rf}. We assume that field $\phi_1$ is subdominant in the background energy density, corresponding to $V_1 \ll V_2$. After $\phi_1$
is annihilated, we assume that not only its energy is transferred into radiation energy density, $\rho_r$,  but also the perturbations carried by $\delta \phi_1$ are transferred into perturbations in radiation, $\delta \rho_r$. We also assume that the perturbations $\delta \rho_r$, before significantly diluted by the background expansion, do not interfere with $\delta \phi_2$ after collision. This can be considered as a higher order back-reaction, perhaps controlled by the ratio $p_1/p_2$, which we take to be small. Also this back-reaction may be negligible as long as the separation between two
stacks of branes are large enough inside the throat. Within these approximations we do not need to follow the perturbations carried by $\delta \phi_1$ before the collision and the perturbations carried by
$\delta \rho_r$ after the collision.  In these approximations we shall follow only the perturbations  $\delta \phi_2$ before and after the collision. This may not be such a bad approximation because physical quantities, such as curvature perturbations, are calculated at the end of inflation when only field $\phi_2$  and its perturbations are present.  Since we do not follow  $\delta \phi_1$ and $\delta \rho_r$, our treatments of perturbations eventually reduce to those of single field model where only  $\delta \phi_2$ is counted before and after the branes collision. However, the effect of $\phi_1$ and radiation is kept in the background so that is how $\delta \phi_2$ feels their presences. 

\subsection{Perturbations Equations}

With these approximations we start the analysis of perturbations in our set up. The metric perturbations in the conformal Newtonian gauge where there is no anisotropy in stress energy tensor  is
\ba
ds^2=-(1+2 \Phi)dt^2+(1-2\Phi)a^2(t)\delta_{ij}dx^idx^j \, ,
\ea
where $\Phi(t, x)$ is the Bardeen potential which in this gauge coincides with the Newtonian potential. 


Defining the  curvature perturbations ${\cal R}_2$ on comoving surface 
$\phi_2 =constant$ via
 \ba
 \label{R-eq}
 \mathcal{R}_2\equiv
H\frac{\dep_2}{\dot{\phi_I}}+\Phi \,  ,
\ea
and introducing the Sasaki-Mukhanov variables $v_2$
 \ba
 \label{z-def}
  v_2=z_2 {\cal R}_2 \quad , \quad  
  z_2  \equiv \frac{a\dot{\phi}_2\gamma_2^{3/2}}{H} \, ,
\ea
one obtains the following standard second order differential equation \cite{Garriga:1999vw}
 \ba
 \label{v eq}
v_2^{''}+ \left( k^2 c_{s }^2 -\frac{z_2''}{z_2} \right) v_2=0  \, .
\ea
Here the prime denotes the derivative with respect to the conformal time $d \tau = d t/a(t)$ and
$c_{s}$ is the sound speed associated with $\phi_2$ perturbations defined by
\ba
c_{s} = \gamma_{2}^{-1} = \sqrt{1-f_2\dot{\phi}_2^2} \, .
\ea
As mentioned before, we work in the ultra relativistic limit where $\gamma_I \gg1$ corresponding to a very small sound speed $c_{s } \ll 1$. This has interesting consequences for non-Gaussianities \cite{Alishahiha:2004eh, Chen:2006nt}.

To solve Eq. (\ref{v eq}) we need to find the form of $z_2$. Plugging different components of 
$z_2$ from the background specified in the previous section, one obtains 
\ba 
 z_2^\pm \propto t^{{\Lambda_\pm} +2} \propto  
 \tau^{- \frac{  {\Lambda_\pm} +2}{ { \Lambda_\pm}  -1} } \, .
\ea
This in turn yields
\ba
\frac{{z''_{2\pm}}}{z_{2\pm}}= \frac{2 \Lambda_\pm^2 + 5 {\Lambda_\pm} + 2}{\left( \Lambda_\pm^2 -2 {\Lambda_\pm} + 1 \right) \tau^2} \simeq  \left( 2 + \frac{9}{{\Lambda_\pm}} \right) {\tau^{-2}} \, ,
\ea
where to get the final approximation the relation ${\Lambda_\pm} \gg 1$ is used as explained in previous section. Consequently, one can cast Eq. (\ref{v eq}) into more conventional form 
 \ba
 \label{v2-eq}
 {v^{''}_2}_{\mp}+\left(k^2{c_s^2}-\frac{\mu_{\mp}^2-1/4}{\tau^2} \right){v_2}_{\mp}=0 \, ,
 \ea
 where  the  indices $\mu_\pm$ are given by
 \ba
 \label{mu-index}
 \mu_\pm \simeq \frac{3}{2} + \frac{3}{{\Lambda_\pm}}  \equiv 
 \frac{3}{2} + 3 \epsilon_\pm \, .
 \ea
 We have defined the ``slow-roll'' parameter  
 $\epsilon_\pm \equiv \frac{-\dot H_\pm}{H_\pm} \simeq \frac{1}{\Lambda_\pm}$ for the later convenience. As argued before, $\epsilon_\pm \simeq 1/N_T \ll 1$.
 
One should solve Eq. (\ref{v2-eq}) for both $t< t_c$ and $t> t_c$ and glue the solutions via the matching conditions. In the limit where we can neglect the running of $c_s$ the solutions of Eq. (\ref{v2-eq})  as usual are given in terms of Hankel functions $H_{\mu_\pm}^{(1)} (- c_s k \tau)$ and $H_{\mu_\pm}^{(2)}(-c_s k \tau)$. At the early stage of inflation when all physically relevant modes are inside the sound horizon, corresponding to $- c_s k \tau \rightarrow \infty$,  the solutions are given in terms of positive frequency modes $\frac{1}{\sqrt{2 c_s k}}e^{- i c_s k\tau}$ created from vacuum. 
This initial condition eliminates $H_{\mu_\pm}^{(2)}(-c_s k \tau)$ and the solution 
of Eq. (\ref{v2-eq}) during the first inflationary stage is
 \ba
 \label{v-}
 {v_2}_{-}=\frac{\sqrt{-\pi\tau}}{2}e^{i\pi(\mu_{-}+1/2)/2}H^{(1)}_{\mu_{-}}(-c_sk\tau) \, .
  \ea

After the collision, due to particles creation, the system is not in vacuum and both solutions 
$H_{\mu_\pm}^{(1)} (- c_s k \tau)$ and $H_{\mu_\pm}^{(2)}(-c_s k \tau)$ are permitted. We parametrize the solution after the collision in terms of the coefficients $\alpha$ and $\beta$ via
\ba
\label{v+}
{v_2}_{+}=\frac{\sqrt{-\pi\tau}}{2}e^{i\pi(\mu_{+}+1/2)/2} \left[ \alpha H^{(1)}_{\mu_{+}}(-c_sk\tau)+\beta  H^{(2)}_{\mu_{+}}(-c_sk\tau) \right] \, .
\ea
In the absence of any brane collision and particles creation $\alpha=1, \beta=0$,  $\mu_- = \mu_+$ and the two solutions (\ref{v-}) and (\ref{v+}) coincides. Our job below is to find the coefficients $\alpha$ and $\beta$ and read off the final value of curvature perturbations as a function of $\alpha$ and $\beta$. For this purpose, we need to impose our matching prescriptions, joining Eqs. (\ref{v-}) and (\ref{v+}) at the time of collision $\tau = \tau_c$.

\subsection{Matching Conditions}

Imposing the matching condition is the most important step in our analysis of perturbations. There are standard prescriptions in literature for imposing the cosmological matching conditions in systems  where there is a sudden jump in equation of state 
\cite{Deruelle:1995kd, Martin:1997zd}. More specifically, in models where the matching conditions are imposed on a comoving hyper-surface one requires that both the extrinsic and the intrinsic curvatures to be continuos across this hyper-surface. These result in the following matching conditions \cite{Zaballa:2009xb, Lyth:2005ze, Zaballa:2006kh}
\ba
\label{match0}
[\Phi]_-^+ = [ {\cal R} ]_-^+ =0 \, ,
\ea
where $\Phi$ is the gauge invariant Bardeen potential and ${\cal R}$ is the comoving curvature perturbations.  

In principle we can also impose these matching conditions in our system. However, as mentioned at the start of this section, we do not have a theoretical control on the dynamics of tachyon condensation, fields annihilation and particles creation. These shortcomings in turn result in a lack of knowledge of $\Phi$ and ${\cal R}$ after the collision. To bypass these shortcomings we have chosen the phenomenological approach where only the perturbations of $\phi_2$ are followed before and after the collision. In this approximation our model is effectively a single field scenario with the extra modification that the effects of  particles creation are kept at the background level. This in turn results in different indices for Hankel functions, $\mu_- \neq \mu_+$. In this view, our method is very similar to  the method used in \cite{Joy:2007na} where the effects of phase transitions in a multiple field model is translated into a sudden violation of slow-roll parameters in an effective single field model. Therefore in our studies here, as in \cite{Joy:2007na}, one expects that the matching conditions (\ref{match0}) are simplified into
\ba
\label{match1}
[v_2]_-^+ = [v_2']_-^+ =0 \, ,
\ea
i.e. both $v_2$ and $v_2'$ are continuous across the surface of branes collision. 

Using the matching conditions (\ref{match1}) in Eqs. (\ref{v-}) and (\ref{v+}) one obtains
\ba 
\label{alpha-beta0}
\alpha&=&-\frac{i\pi x}{4}e^{i\delta} \left[H'^{(1)}_{\mu^{-}}(x)H^{(2)}_{\mu^{+}}(x)
-H^{(1)}_{\mu^{-}}(x)H'^{(2)}_{\mu^{+}}(x) \right]  \nonumber\\
\beta&=&\frac{i\pi
x}{4}e^{i\delta}\left[ H'^{(1)}_{\mu^{-}}(x)H^{(1)}_{\mu^{+}}(x)
-H^{(1)}_{\mu^{-}}(x)H'^{(1)}_{\mu^{+}}(x) \right] \, ,
\ea
where the derivatives are with respect to the arguments of the Hankel functions and 
\ba
x \equiv - c_s k \tau_c \quad , \quad \delta \equiv \frac{\pi}{2} ( \mu_{-} - \mu_{+} ) \, .
\ea

\section{Power Spectrum}
\label{power}
We are in a position to calculate the curvature power spectrum ${\cal P}_{\cal R}$ at the end of inflation.  The curvature power spectrum is defined via
\begin{eqnarray}
\delta^3({\bf k}-{\bf k}^\prime)\mathcal{P}_{\mathcal{R}}=\frac{4\pi
k^3}{(2\pi)^3}\langle\mathcal{R}({\bf k}^\prime)^*\mathcal{R}({\bf k})\rangle \, .
\end{eqnarray}
Since only the field $\phi_2$ is present at the end of inflation we only need the power spectrum of ${\cal R}_2$.  Using Eqs. (\ref{R-eq}) and (\ref{z-def}) one obtains
\ba
{\cal R}_2 = \frac{H v_2}{a \dot \phi_2 \gamma^{3/2}} \, .
\ea
At the later stage of inflation, when the mode of interest is outside of the sound horizon and 
$- c_s k \tau \rightarrow 0$, one can approximate
\begin{eqnarray}
H^{(2)}_{\mu_+}(-c_s k\tau)\simeq - H^{(1)}_{\mu_+}(-c_s k\tau)\simeq
\frac{i}{\pi}\Gamma(\mu_+)\left(- \frac{c_s k\tau}{2}\right)^{-\mu_+}\,.
\end{eqnarray}
Plugging this into the power spectrum definition yields 
\ba
\label{PR}
\mathcal{P}_{\mathcal{R}} = \mathcal{P}_{{\mathcal{R}}_0 } \left|\beta-\alpha  \right|^2\, ,
\ea 
where $\mathcal{P}_{{\mathcal{R}}_0 }$ represents the curvature power spectrum in the absence of branes collision 
\begin{eqnarray}
\label{PR0}
 \mathcal{P}_{{\mathcal{R}}_0 }&\simeq&
\left( \frac{H^2 (1- \epsilon_+) }{2 \pi \dot \phi_2}   \frac{\Gamma(\mu_+)}{\Gamma(3/2)}  \right)^2    \left(\frac{- c_s k  \tau}{2} \right)^{3- 2 \mu_+}  \,.
\end{eqnarray}
 Observationally $\mathcal{P}_{{\mathcal{R}}_0 }$ is fixed by the COBE normalization, 
 $ \mathcal{P}_{{\mathcal{R}}_0 } \simeq 2\times 10^{-9}$, when the mode corresponding to the current Hubble radius leaves the horizon at about 60 e-folds before the end of inflation with
 $k= a H \gamma$. One can use the COBE normalization to fix a combination of parameters. Specifically, using Eqs. (\ref{phi-back}) and (\ref{Lambda+}) we have
\ba
 \mathcal{P}_{{\mathcal{R}}_0 } \simeq  \frac{\Lambda_+^4}{4 \pi^2 \lambda_2}   \, .
\ea
Combined with Eq. (\ref{Lambda+}) the COBE normalization yields $\sqrt \lambda_2 m_2^2/M_P^2 \simeq 2 \times 10^{-3}$. Assuming that $\Lambda_+ \lesssim N_T =60$, this yields $\lambda_2 \sim 10^{14} $. This is the infamous fine-tuning problem associated with the DBI inflation \cite{Alishahiha:2004eh, Baumann:2006cd}.  Correspondingly $m_2/M_P \sim 10^{-6}$. 

To calculate the background curvature perturbations spectral index, $n_{\cal R}^0$, we note that at the time of horizon crossing $d \ln k = d \ln a + d \ln H + d \ln \gamma 
= ( 1- \epsilon_+ + \dot \gamma/\gamma H) \, H \, dt$. Using this identity in (\ref{PR0}) yields 
\ba
\label{nR0}
n_{\cal R}^0-1 \equiv \frac{d \ln  \mathcal{P}_{{\mathcal{R}}_0 } }{d \ln k}
\simeq \frac{2 \left(  \frac{\dot \phi_2}{H^2}\right) }{H \left(1- \epsilon_+ +  \frac{\dot \gamma}{\gamma H} \right)}  \frac{d}{d t} \left( \frac{H^2}{\dot \phi_2}  \right)
\simeq {\cal{O}} (\epsilon_+^2) \, ,
\ea 
where to get the final result the relations  $H \propto t^{-1}$ and $\dot \phi_2 \propto t^{-2}$ 
have been used from the background. This indicates that $n_{\cal R}^0$ is nearly scale invariant up to ${\cal{O}} (\epsilon_+^2)$ \cite{Shandera:2006ax, Bean:2007eh, Bean:2007hc}.

\begin{figure}[t]
\includegraphics[width=8cm]{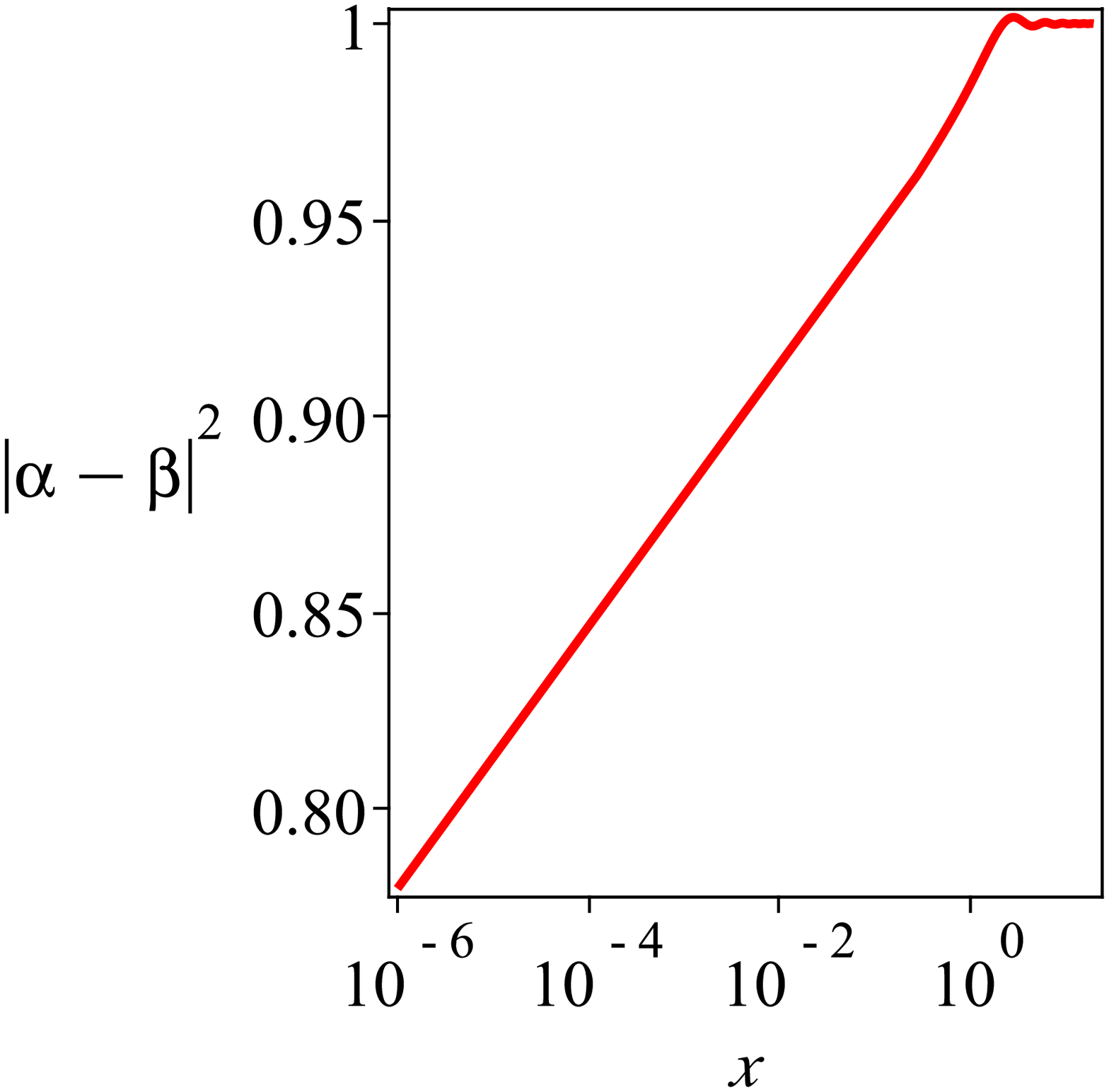} \hspace{0cm}
\includegraphics[width=8cm]{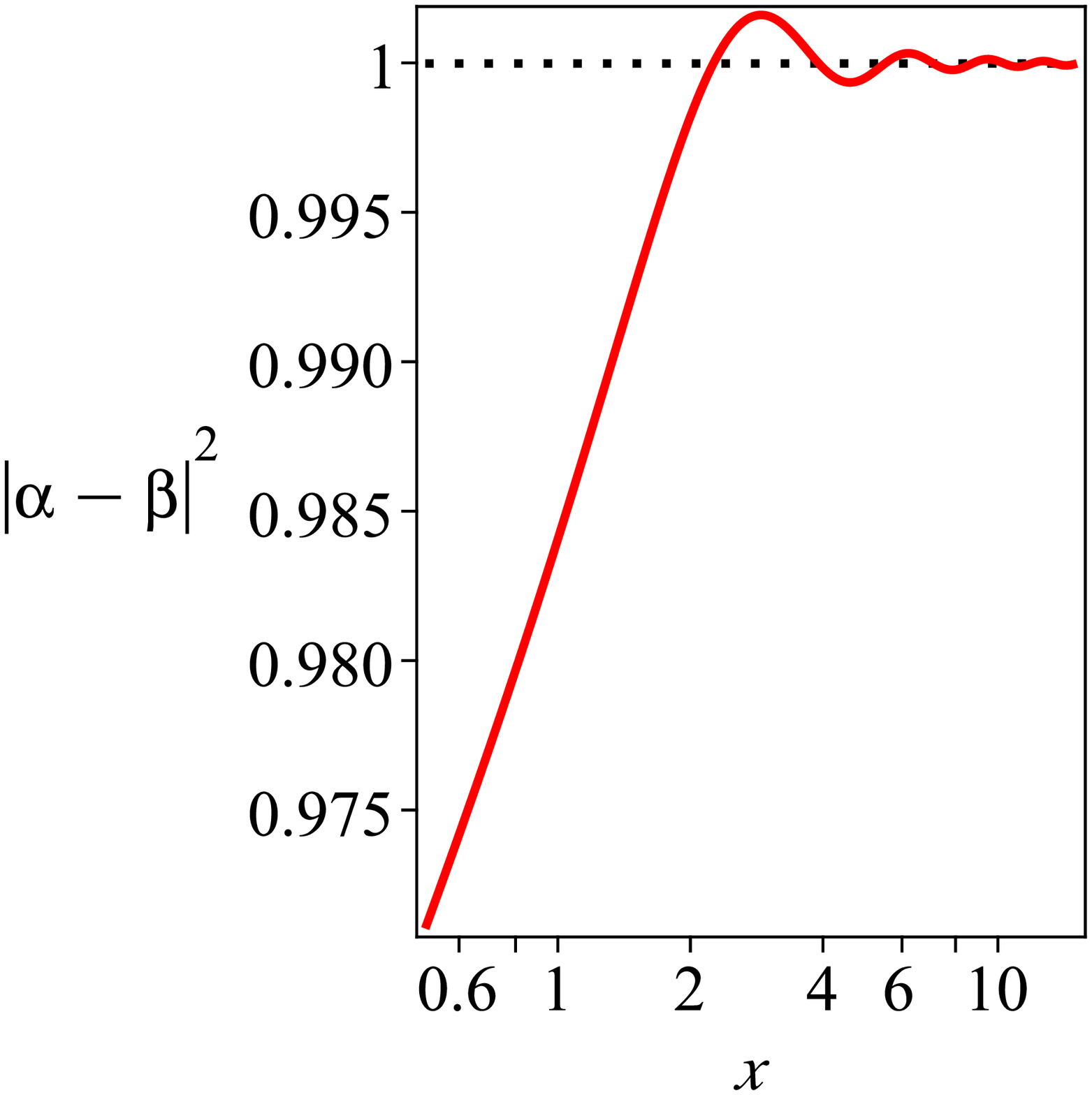}
 \caption{Here we plot the transfer function given by Eq. (\ref{alpha-beta}) with $\Lambda_- = 45$ and $\Lambda_+ = 40$. For large scales, $x\ll 1$, it has a monotonic behavior as indicated by Eq. (\ref{alpha-beta-sup}) whereas for small scales, $x\gg1$, it has an oscillatory behavior approaching the unity. The transition takes place for intermediate scales, $x\simeq 1$, modes which leave the horizon around the time of branes annihilation. 
  }
 \vspace{0.5cm}
 \label{transfer-fig}
\end{figure}

The effects of branes collision and particles creation in the power spectrum Eq. (\ref{PR}) 
are encoded in the transfer function $\left| \beta-\alpha \right|^2$. In the absence of any feature, $\alpha=1, \beta=0$ so we obtain the standard single field DBI inflation \cite{Alishahiha:2004eh}. Below we would like to investigate the behavior of the transfer function $\left| \beta-\alpha \right|^2$ for different modes.  Using the explicit formulae of $\alpha$ and $\beta$ given in Eq. (\ref{alpha-beta0}) one obtains
\ba
\label{alpha-beta}
\left| \beta-\alpha \right|^2 =  \frac{\pi^2 }{4} \left[ \left(  J_{\mu_+} J'_{\mu_-}  -   J'_{\mu_+} J_{\mu_-}   \right)^2 
 + \left(  J_{\mu_+} Y'_{\mu_-}  -   J'_{\mu_+} Y_{\mu_-}   \right)^2  \right] x^2 \,.
\ea
As mentioned before we have defined $x\equiv - c_s k \tau_c = k/k_c$ where $k_c$ represents the critical mode which leaves the sound horizon at the time of branes collision during inflation.
One can check that in the absence of any feature, corresponding to $\mu_-= \mu_+$, one obtains $\left| \beta-\alpha \right|^2 = 1$ as expected. As mentioned previously, the effects of branes annihilation and particles creations are such that  $ \Delta \mu \neq 0$ where  
$\Delta \mu \equiv \mu_+ - \mu_-$. 

The spectral index is also modified in the presence of the transfer function as follows
\ba
\label{nR}
n_{\cal R} -1 = n_{\cal R}^0 -1 + x \frac{d  }{ dx} \left| \beta-\alpha \right|^2 \, .
\ea

Now let us look at the shape of the transfer function for different  length scales. In {\bf Fig.} \ref{transfer-fig} we have plotted the transfer function. As a measure
of the length scales of the perturbations, we note that $x=1$ corresponds to the critical mode
$k= k_c$ which leaves the sound horizon right at the time of branes annihilation. Then depending on whether the mode of interest leaves the sound horizon before the collision or after the collision  one respectively has $x\ll 1$ or $x\gg1$. 

\begin{figure}[t]
\includegraphics[width=8cm]{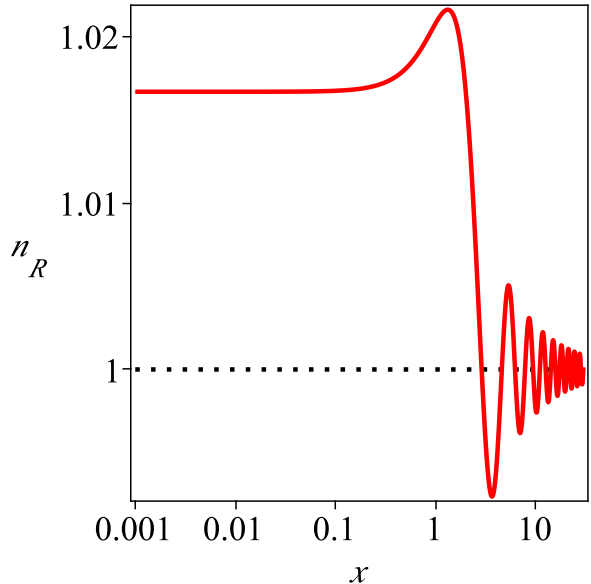} \hspace{0cm}
 \caption{Here we plot the spectral index $n_{\cal R}$. For super-horizon modes, $x\ll 1$, our analytical formula Eq. (\ref{nR}) gives a very accurate prediction for $n_{\cal R}$.   For small scales, $x\gg1$, the oscillatory behavior of $n_{\cal R}$ is clearly seen as suggested by Eq. (\ref{alpha-beta}) where $n_{\cal R}$ approaches its background value $n_{\cal R}^0$  denoted by the dotted line. The transition takes place around $x\simeq 1$ for modes which leave the horizon around the time of branes annihilation. The parameters are as in {\bf Fig.} \ref{transfer-fig}}.
 \vspace{0.5cm}
 \label{nR-fig}
\end{figure}

First consider the super-horizon scales, $x\ll 1$. This corresponds to modes which leave the sound horizon before the collision.   Using the small argument limit  of the Bessel functions we obtain
\ba
\label{alpha-beta-sup}
\left| \beta-\alpha \right|^2 \simeq 
 \frac{\Gamma^2(\mu_-)}{ 4 \Gamma^2(\mu_+ +1)}    \left(  \mu_-+ \mu_+      \right)^2 \left(\frac{x}{2} \right)^{2 \Delta \mu  } \, .
\ea
Our background is such that $\Lambda_+ < {\Lambda_-}$ due to branes collision and energy transfer into radiation so $\Delta \mu>0$. As a result the transfer function scales 
monotonically for super-horizon modes.  Correspondingly, the change in spectral index is obtained to be
\ba
\label{nR-sup}
\Delta  n_{\cal R} \equiv     n_{\cal R} -  n_{\cal R}^0 \simeq  2 \Delta \mu \, .
\ea
We see from Eq. (\ref{nR-sup})  that the spectral index for the 
super-horizon scales become more blue-tilted as compared to the background  spectral index. This can also be seen in {\bf Fig.} \ref{nR-fig}.

Now consider the sub-horizon modes, corresponding to $x \gg1$. These are the modes which are deep inside the sound horizon at the time of branes collision. Physically one expects these modes to live in their local Minkowski background and should not be affected by branes collision and particles creation. Indeed, using the large argument limit of the Bessel functions 
$J_\nu(z) \simeq   \sqrt{\frac{2}{\pi z}}    \cos\left(z- (\nu+\frac{1}{2}) \frac{\pi}{2} \right) $ and 
$Y_\nu(z) \simeq   \sqrt{\frac{2}{\pi z}}    \sin\left(z- (\nu+\frac{1}{2}) \frac{\pi}{2} \right) $ for $z \gg 1$
we find
\ba
\label{alpha-beta-sub}
\left| \beta-\alpha \right|^2 \simeq 
1+ {\cal O} (x^{-2})  \, .
\ea
As expected the transfer function approaches unity so $n_{\cal R} \simeq  n_{\cal R}^0$. 
However, both the transfer function and the spectral index are superimposed by oscillatory modulations with decaying amplitude \cite{Contaldi:2003zv, Romano:2008rr, Zarei:2008nr}.

The most interesting effects of branes annihilation and particles creation is on the intermediate scales, $x\simeq1$. These are the modes which leave the sound horizon around the time of branes collision. In Figures \ref{transfer-fig} and \ref{nR-fig} we have plotted the transfer function and the spectral index. As can be seen, for super-horizon scales the transfer function is monotonically increasing whereas for small scales it oscillates around unity. Correspondingly, the spectral index for super-horizon scales is well approximated by Eq. (\ref{nR-sup}) whereas for small scales it oscillates around $n_{\cal R}^0 $. The non-trivial transition, happening at the intermediate scales, can be seen in both figures.  

As discussed before, the effects of branes annihilation and particles creation result in 
 $\Delta \mu \neq 0$. So far in the perturbations analysis  we kept $\mu_\pm$ and $\Lambda_\pm$ undetermined. Now let us specify  $\Delta \mu$ in terms of our background model parameters. Starting with Eqs.  (\ref{lambda-}), (\ref{Lambda+})  and (\ref{mu-index}) we obtain
 \ba
 \Delta \mu = \mu_+ - \mu_- &\simeq& \frac{3}{\Lambda_+} - \frac{3}{ {\Lambda_-}} \nonumber\\
 &\simeq&\frac{3}{2 \Lambda_+} \frac{p_1}{p_2} \frac{m_1^2}{m_2^2} \, .
 \ea
 To get the final result we used the approximation that $\Lambda_+ \simeq  {\Lambda_-}$. Physically, this is motivated from the assumption that the energy density
 associated with stack 1 is subdominant compared to the energy density stored in stack 2 so the energy transferred into radiation is much smaller than the background potential energy density.  This assumption is necessary for the consistency of our matching condition. We have neglected the interference of perturbations $\delta \phi_1$ and $\delta \rho_r$ on the perturbations $\delta \phi_2$. As a consequence, we have imposed the simple matching condition (\ref{match1}). For this approximation to be valid, we require that $\rho_r/V(\phi_2) \ll1$ or equivalently  
 $({\Lambda_+}- \Lambda_-) /{\Lambda_+} \ll1$.
 
We do not have good theoretical control on the masses $m_1$ and $m_2$ which were originated from the back-reactions of Kahler moduli stabilization and background fluxes on the mobile branes. As a simple ansatz, we may take $m_1 \simeq m_2$. This corresponds to the case that
the back-reactions imposed from the background compactification on a single brane is  independent of its position inside the throat. This may not be an unreasonable assumption. With this approximation, the consistency of our approximation requires $p_1\ll p_2$. Also we are working in the probe brane approximations so the mobile stacks of branes should not deform the background AdS geometry significantly. For this to be the case we also need to impose $p_1+ p_2 \ll N$. As an example, taking $\Lambda_+ =40, p_1=1$ and $ p_2 =7$ results in  
$ \Delta  n_{\cal R} \simeq 0.01$. 

One may also compare $\Delta  n_{\cal R}$ induced from branes collision 
to the  ${\cal{O}} (\epsilon_+^2) $ contributions to  the background spectral index in Eq. (\ref{nR0}).  We have $\Delta \mu \simeq 3 (\Lambda_- - \Lambda_+)/\Lambda_+^2 \simeq  
3 (\Lambda_- - \Lambda_+) \epsilon_+^2$. For a reasonable value of $(\Lambda_- - \Lambda_+) =5$ which was used in our numerical examples in Figures \ref{transfer-fig} and \ref{nR-fig} we have $\Delta \mu \sim 15 \epsilon_+^2$ or $\Delta  n_{\cal R} \sim 30 
\epsilon_+^2$. This indicates that the contributions of the branes collision to the spectral index is much bigger than the sub-leading ${\cal{O}} (\epsilon_+^2) $ contributions to the background $ n_{\cal R}^0$. 

Observationally the result $\Delta  n_{\cal R} >0$ is disfavored \cite{Komatsu:2010fb} noting that the background spectral index is already nearly scale invariant as seen in Eq. (\ref{nR0}). To remedy this problem, it may be necessary to embed the simple model of ultra relativistic DBI inflation here to a generic setup of brane inflation where a period of slow-roll brane inflation takes place before DBI inflation starts. This can help to obtain a red spectral index for large scales whereas for smaller scales the spectral index can be blue as we have here.

\section{Discussions and Conclusions}
\label{conclusion}
In this work we have studied the effects of branes annihilation and particles creation in models of DBI inflation where the branes are moving ultra relativistically. In a typical string theory compactification the process of branes and anti-branes annihilation is a generic phenomenon. It is therefore an interesting question how these collisions can affect the dynamics of inflation which may also be detectable in the sky.

The process of branes annihilation is determined by the dynamics of open string tachyon condensation. In a complicated background, such as the warped throat with background fluxes, this is a non-trivial process. The time-scale of tachyon condensation and branes annihilation is given by the inverse of the warped string mass scale, $m_s h(r_0)^{1/4}$. In our effective field theory approach the Hubble expansion rate is considerably smaller than this mass scale so for practical purposes we can take the process of branes annihilation and particles creation to be instantaneous. In our set up the first inflationary stage is a two-field DBI inflation model.
After the first stack of branes is annihilated with the background anti-branes a burst of closed strings modes are created. These particles are rapidly diluted by inflation and we eventually end up with a single field DBI inflation system. In order to connect the perturbations in the second inflationary stage to the perturbations generated before the collision we have to use proper matching conditions. Here are the major simplifications imposed in our analysis. 
It is assumed that the energy in field $\phi_1$, which is annihilated during inflation, is sub-dominant compared to energy stored in field $\phi_2$ which carries the second stage of inflation. It is assumed that not only the background energy density in $V_1(\phi_1)$ is transferred into radiation $\rho_r$ but also the perturbations carried by $\delta \phi_1$ are also transferred into  perturbations in radiation, $\delta \rho_r$. With these simplifications we can follow the perturbations in $\phi_2$
before and after the collision using the simple matching conditions (\ref{match1}).
Although we did not take into account the perturbations $\delta \phi_1$ and $\delta \rho_r$ in our analysis but the effects of branes annihilation and particles creation are present at the level of background. In our analysis this translated into $\Delta \mu$ to be non-zero.  Interestingly, this is very similar to model  studied in  \cite{Joy:2007na, Joy:2008qd} where the feature is due to a sudden change in slow-roll parameters. In  \cite{Joy:2007na, Joy:2008qd} the sudden change in slow-roll parameters was motivated from a phase transition \cite{Abolhasani:2010kn} during hybrid inflation model  \cite{Linde:1993cn, Copeland:1994vg, GarciaBellido:1996qt}. It is very interesting that our model, motivated from a completely different setup,  shares the same results.  

We have calculated the curvature perturbations power spectrum. The effects of branes annihilation are encoded in the transfer function $\left| \beta-\alpha \right|^2$. For super-horizon scales, modes which are outside the sound horizon at the time of collision, the transfer function scales monotonically. Consequently, the spectral index is more blue compared to the background spectral index.  On the other hand, for small scales, modes which are inside the sound horizon at the time of collision, the transfer function converges to unity with decaying superimposed oscillatory modulations. The non-trivial transitions in transfer function happen for modes which leave the sound horizon around the time of collision.
The superimposed oscillatory modulations can have interesting observational consequences on
CMB analysis \cite{Joy:2008qd, Flauger:2009ab, Biswas:2010si}. Furthermore, there are strong constraints from non-Gaussianity bounds on standard DBI inflation  \cite{Shandera:2006ax, Bean:2007eh, Bean:2007hc}. It would be interesting to calculate the level of non-Gaussianities produced
in our setup to put further constraints on the model parameters. 

Our results here should be compared with the results obtained in \cite{Battefeld:2010rf} where similar problem was studied but in the context of slow-roll brane inflation. In \cite{Battefeld:2010rf}, like here, it was assumed that the perturbations $\delta \phi_1$ are transferred into 
$\delta \rho_r$.  However,  in \cite{Battefeld:2010rf}  the matching conditions (\ref{match0}) have been used instead of the simple matching conditions (\ref{match1}). As a result, the transfer function  for large and intermediate scales  in \cite{Battefeld:2010rf} share the same behavior as in our 
case here. However,  in \cite{Battefeld:2010rf} (see also \cite{Flauger:2009ab, Biswas:2010si, Flauger:2010ja}), the transfer function for small scales has a constant oscillatory modulation whereas in our case the amplitude of oscillations decays for these modes.  For physical reasons, one expects that small scales should be blind to the process of branes annihilation and particles creation so the transfer function is expected to approach unity for these modes. To remedy this problem  several methods were put forward in \cite{Battefeld:2010rf}. Our approach to use the matching conditions (\ref{match1}) may be interpreted as a pragmatic approach towards the methods speculated in   \cite{Battefeld:2010rf}. 

To simplify our analysis  we considered the model where the branes are moving ultra relativistically. In practice one expects that a successful  brane inflation model consists both stages of slow-roll and fast-roll brane inflation. It is possible that during early stage of inflation the branes are moving slowly so one may get a few dozens of e-foldings during the slow-roll  regime \cite{Kachru:2003sx}. As the branes are moving towards the bottom of the throat they reach the speed limit and one should use the DBI inflation methods as we used
in this work. The process of branes annihilation and particles creation can happen either during slow-roll regime, as in \cite{Battefeld:2010rf}, during the ultra relativistic fast-roll limit, as in our case here, or in between these two stages. In the third case, one can not perform analytical studies and a numerical investigation is necessary. It would be interesting to study the effects of branes annihilation in a generic model of brane inflation containing both stages of slow-roll inflation and DBI inflation.


\section*{ Acknowledgments}

We would like to thank F. Arroja,  D. Battefeld, T. Battefeld, N. Khosravi, 
L. McAllister, M.H. Namjoo, M. Sasaki and G. Shiu for useful discussions and correspondences.  H.F. would like to thank  Yukawa Institute for Theoretical Physics (YITP) for the hospitalities during  the activities ``Gravity and Cosmology 2010" and ``YKIS2010 Symposium: Cosmology -- The Next Generation"  when this work was in progress. 

\vspace{1cm}

\end{document}